\documentclass[aps,prl,floatfix,twocolumn,tightenlines,amsmath,amssymb]{revtex4-1}

\usepackage{graphicx}
\usepackage{amssymb}
\usepackage{color}
\usepackage{psfrag}
\usepackage{ifsym}
\usepackage{epstopdf}
\usepackage{soul}
\usepackage{tabularx}
\usepackage[latin1]{inputenc}
\usepackage{amsmath}
\usepackage{color}

\newcommand{\beq}{\begin{equation}}
\newcommand{\eeq}{\end{equation}}

\begin{document}

\title{Frequency cavity pulling induced by a single semiconductor quantum dot.}

\author{Daniel Valente$^{2,4}$}
\author{Jan Suf\mbox{}fczy\'nski$^{1,3}$}
\author{Tomasz Jakubczyk$^{3}$}
\author{Adrien Dousse$^{1}$}
\author{Aristide Lema\^itre$^{1}$}
\author{Isabelle Sagnes $^{1}$}
\author{Lo\"ic Lanco$^{1}$}
\author{Paul Voisin$^{1}$}
\author{Alexia Auff\`eves$^{2}$}\email{alexia.auffeves@grenoble.cnrs.fr}
\author{Pascale Senellart$^{1}$}\email{pascale.senellart@lpn.cnrs.fr}


\affiliation{$^{1}$ CNRS-LPN Laboratoire de Photonique et de Nanostructures, Route de Nozay,  91460 Marcoussis, Francee}

\affiliation{$^{2}$ CEA/CNRS/UJF joint team ``Nanophysics and Semiconductors'', Institut N\'eel-CNRS, BP 166, 25 rue des Martyrs, 38042 Grenoble Cedex 9, France}

\affiliation{$^{3}$ Institute of Experimental Physics, Faculty of Physics, University of Warsaw, Ho$\dot{z}$a 69, PL-00-681 Warszawa, Poland}

\affiliation{$^{4}$ Instituto de F\'isica, Universidade Federal de Mato Grosso, Cuiab\'a MT, Brazil}

\begin{abstract}

We investigate the emission properties of a single semiconductor quantum dot deterministically coupled to a confined optical mode in the weak coupling regime. A strong pulling, broadening and narrowing of the cavity mode emission is evidenced when changing the spectral detuning between the emitter and the cavity. These features are theoretically accounted for by considering the phonon assisted emission of the quantum dot transition. These observations highlight a new situation for cavity quantum electrodynamics involving spectrally broad emitters.

\end{abstract}

\maketitle

Progresses in nanotechnologies allow to efficiently couple  solid-state optical resonators such as dielectric cavities or plasmonic antennas to artificial atoms like NV centers in diamond \cite{bensonplasmon}, nanocrystals \cite{colloidal} or semi-conductor quantum dots \cite{pelton2002,doussenat,badolato}. Such optical media benefit from strong light-matter interaction and show single photon sensitivity \cite{MichlerScience2000,michlernature2000,reinhard,bose,loo2012},
opening the path to scalable quantum information processing on chip. However, despite their similarities with real atoms, artificial ones show very different emission properties, due to their coupling to the solid-state environment they are embedded in, whether
phononic \cite{phononNV,besombes,peter} or electrostatic \cite{augerQDs,berthelot, poizat}.  This strong interaction leads to a broadening of the emission line which can eventually overcome the cavity line.

Although this "bad emitter regime"   has been mostly considered a drawback for  practical purposes, recent works have suggested that it could be a resource for quantum light generation. Indeed, coupling a spectrally broadened emitter to a high quality factor resonator induces Purcell enhancement of mechanisms resonant with the mode, leading to the intense emission of photons at the resonator frequency, even when the detuning between the mode and the emitter is large. This "cavity-feeding" effect has been extensively studied for semiconductor quantum dots (QDs) \cite{kaniber,suff,ates,volz} and has been proposed to build wavelength stabilized single photon sources \cite{auffeves} or single emitter based lasers \cite{auffeves2}. Very recently, these ideas have been exploited to demonstrate a narrowband tunable single photon source at room temperature with a NV center coupled to an optical fiber based cavity \cite{becher}. 

From the fundamental point of view, this bad emitter limit allows exploring a situation for cavity quantum electrodynamics not investigated with real atoms so far. In this framework, semiconductor self-assembled QDs are particularly interesting objects. Recent progresses in technology have allowed to prepare the ideal system of a single QD presenting a controlled coupling with a single mode optical resonator \cite{badolato,dousse08}.   
In the present work, we study the emission of a QD deterministically coupled to a pillar cavity mode of low quality factor and evidence a new effect appearing in this bad emitter regime. Phonon assisted emission leads to a pulling of the cavity emission toward the QD line as well as broadening or narrowing depending on the QD-cavity detuning. These observations present similarities with the well-known cavity pulling effect in lasers with thousands of atoms \cite{bennett1,bennett2,siegman, casperson}. They are here attained with a single emitter in the weak coupling regime.

\color{black}

We study the emission of single InGaAs QDs inserted in GaAs/AlGaAs micropillar cavities and present experimental results on two QD-pillars  cavities, both operating in the weak coupling regime. The first pillar has been obtained without a deterministic technology: it embeds several QDs including QD1 close to spectral resonance with the cavity mode but  not located at the maximum of the electric field. The  second pillar has been deterministically fabricated using the in-situ lithography technique. It embeds QD2, close to spectral resonance with the cavity mode, located within 50 nm from its center, ensuring an optimal Purcell effect (see reference  \cite{dousse08} for details on the pillars fabrication). 
Each pillar is non-resonantly excited in the continuous wave regime at 800 nm. The photoluminescence signal is collected from the top of the pillar and analyzed with a spectrometer with a 50-100 $\mu $eV spectral resolution. 

\begin{figure}[h!]\begin{center}\includegraphics[width=1.00\linewidth]{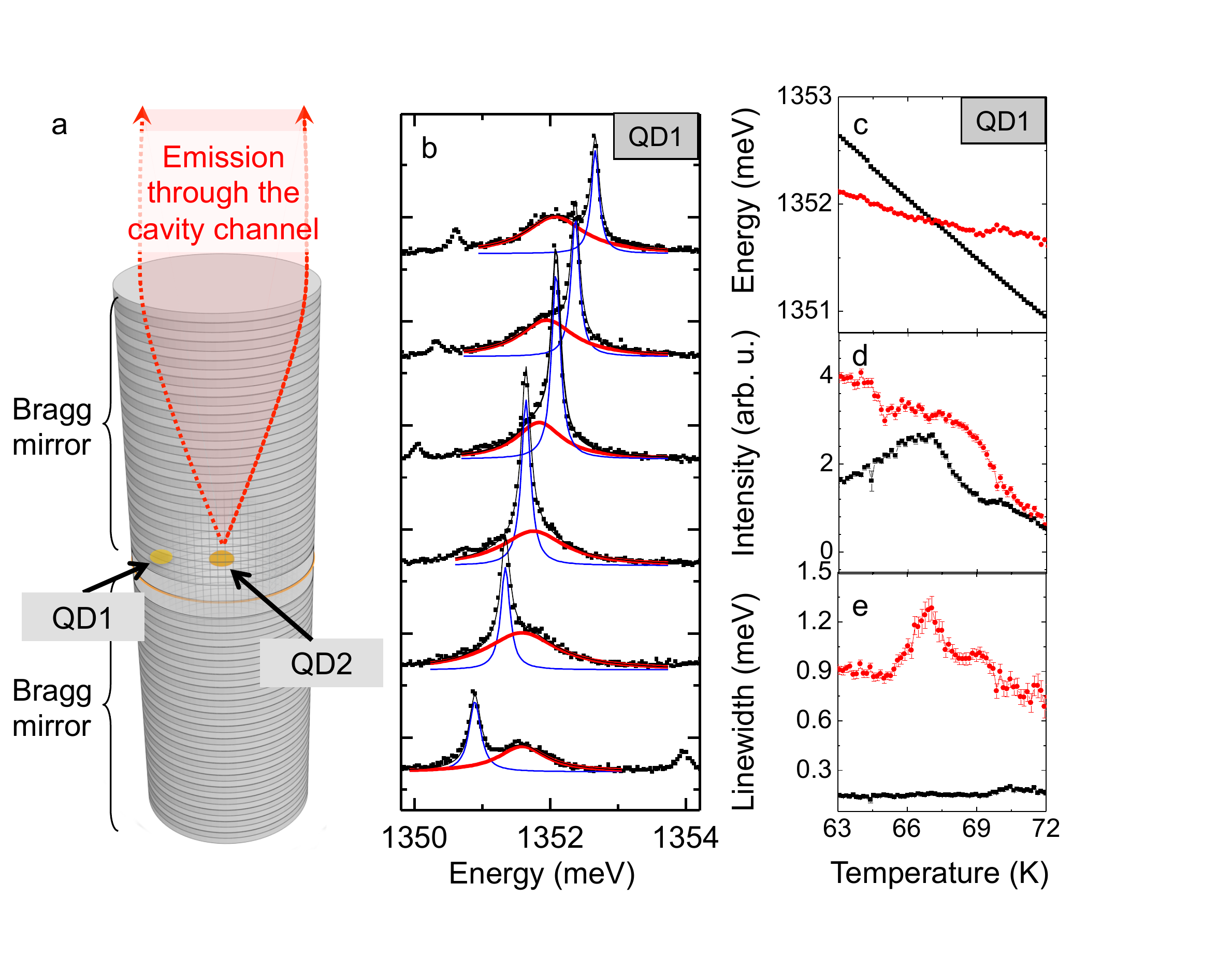}\caption{a: System under investigation: a single QD is coupled to the confined optical mode of a micropillar cavity. The emission from the cavity channel is measured. We study two pillars: the first one embeds QD1 non deterministically coupled to the mode, whereas the second one is deterministically and optimally coupled to QD2.  Both pillars operate in the weak coupling regime.  b: Emission spectra measured for various temperatures on QD1, presenting a poor coupling to the pillar mode. The two narrow emission lines corresponds to the single QD emission lines and the broad emission to the cavity mode.  The red and blue lines are lorentzian fits to the experimental data. c-e: Characteristics of the QD and cavity system deduced from emission spectroscopy: Energy (c), Intensity (d) and Linewidth (e) of the QD (black) emission line and the cavity (red) emission line.  }\label{fig:2}\end{center}\end{figure}

Figure 1b shows the emission spectra  as a function of temperature and energy for the cavity non deterministically coupled to QD1. Spectrally narrow emission lines with a strong energy dependence on temperature arise from the QD emission. The broader line presenting a smaller energy variation with temperature corresponds to the emission within the cavity mode. This emission observed even when the detuning between the QD and the mode is large is  the so-called "cavity feeding"\cite{kaniber,suff}, namely the Purcell enhancement of broadband "background" emission. When several QDs are inserted in the cavity without deterministic positioning, this emission partially arises from non resonant QDs \cite{kaniber,ates} which behave as internal white sources and allow to measure the intrinsic characteristics of the mode. The properties of the QD1-cavity system  are extracted by routinely fitting the experimental curves with a sum of two lorentzian peaks, one for the QD lines (blue),  the other for the cavity emission line (red). The line positions, intensities and linewidths  are shown as a function of temperature in figure 1c. The QD and cavity emission lines  continuously shift to lower energies with temperature. The QD emission slightly increases at resonance because of a small Purcell effect while the cavity line intensity  does not show changes related to the QD resonance. Last, the cavity linewidth slightly varies across the resonance, but only in the range of 20$\%$. Note that in all experiments reported here, the measured QD emission linewidth  is limited by the  experimental setup resolution.
  
Figure 2 reports significantly different features, observed on the deterministically coupled QD2-cavity device.  Figure 2a shows the emission intensity as a function of energy and temperature for QD2 where three   emission lines (charged exciton (CX),  neutral exciton  (X) and  biexciton (XX)) are successively brought into resonance with the cavity mode through temperature tuning.  Figure 2b and 2c present two emission spectra for two  detunings (vertical white lines in figure 2a) where a cavity emission linewidth differing by almost a factor 2 is observed,  showing a strong dependence of the mode properties on the QD transitions-cavity detuning.

This dependence is confirmed in figure 2d-f presenting the emission properties deduced by fitting each spectrum by a sum of lorentzian peaks.  
 We only plot the fit results for detunings where the two lines are well resolved spectrally. Unlike in Fig. 1c,  the mode energy displays a S-like temperature dependence (figure 2d) in between two spectral resonance between the QD transition line.
Figure 2e presents the spectrally integrated intensity emission of the QD and cavity lines. For each spectral resonance, the  increase in the QD emission intensity corresponds to a significant decrease of the mode intensity. Finally, the cavity emission linewidth shown in figure 2f presents strong variations with detuning, exceeding a factor 2.5: the cavity line is  broader when the mode stands in between two QD lines.

\begin{figure*}[t]\begin{center}\includegraphics[width=1\linewidth]{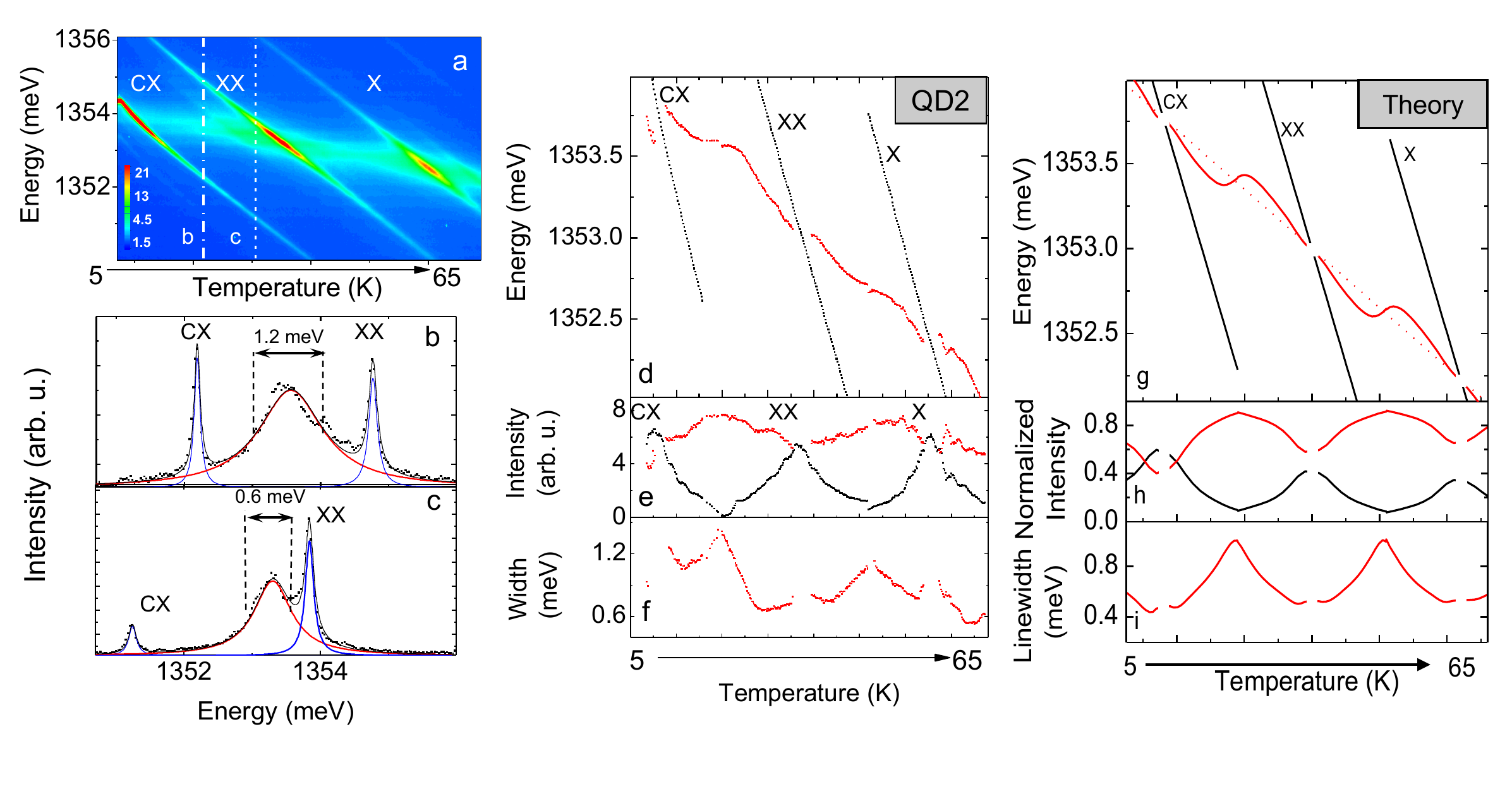}\caption{a: Emission intensity as a function of energy and temperature for the pillar coupled to QD2. b,c: emission spectra extracted from the temperature flow presented in a, corresponding to the two vertical white lines. Black symbols: experimental points. Black lines: fit with the sum of three lorentzian peaks. Red line: cavity peak. Blue lines: CX and XX peaks. d-f: Characteristics of the QD2-pillar device deduced by fitting the spectra with sums of lorentzian peaks: Energy (top), Intensity (middle) and linewidth (bottom) of the QD (black) emission line and the cavity (red) emission lines. The QD linewidth which is below the spectral resolution for all temperature is not shown.  g: Calculated energies of the apparent QD  (black) and  mode (red) lines as a function of temperature. The red dotted line in the  bare cavity mode energy.  h: Normalized emission intensity of the QD line (black) and the mode line (red) as a function of temperature. i: Calculated width of the apparent cavity line as a function of temperature.   }\label{fig:3}\end{center}\end{figure*}

Observing these new features has been made possible thanks to a full control in the QD-cavity coupling:   the cavity distortion is observed on every deterministically coupled QD-pillar device \cite{dousse08} with quality factor around 1000-5000, the lower the Q, the larger the  distortion.  In an ideally coupled QD-cavity system, the cavity emission is expected to arise mostly from phonon sidebands in the small detuning range \cite{rae}.  
 In the following, the emission spectra of a single QD exciton weakly coupled to a cavity mode and to the phonon bath continuum \cite{rae,savona} is calculated using  Fermi's golden rule. Contrary to recent works \cite{hughes,vallee}, no direct incoherent pumping of the mode is introduced: we show that the cavity distortion indeed  arises from cavity pulling effects when the QD phonon sidebands mainly participate to cavity feeding. 
 
 The coupling of a QD discrete state to acoustic phonons leads to the appearance of emission sidebands which spread over a few meV spectral range \cite{besombes,peter}. They result from the emission of a photon assisted by acoustic phonon emission or absorption. The initial state is the excited quantum dot with a given thermal distribution for the acoustic phonon bath. It decays towards a continuum of electromagnetic and phononic modes. The emission spectrum $S_{\mathrm{QD}}(\omega,T)$ of a QD in bulk at temperature T  is analytically calculated by exact diagonalization of the independent boson model Hamiltonian (see supplementary materials). At low temperatures, the spontaneous emission spectrum is given by $S_{\mathrm{QD}}(\omega,T) =\Gamma_{\mathrm{ZPL}}(\omega)+\Gamma_{\mathrm{1PL}}(T,\omega)$. The first term $\Gamma_{\mathrm{ZPL}}(\omega)$ describes the QD zero-phonon line (ZPL),  proportional to a Dirac delta function, which is phenomenologically substituted here by a lorentzian. 
The second term accounts for the temperature dependent one phonon line (1PL), given by $\Gamma_{\mathrm{1PL}}(T,\omega) = (n(\omega,T)+1) \Gamma_{\mathrm{1PL}}(\omega)+n(\omega,T) \tilde{\Gamma}_{\mathrm{1PL}}(\omega)$, where the first term corresponds to the emission of  a phonon and the second term to absorption. $\Gamma_{\mathrm{1PL}}(\omega)$ is the one phonon line at zero temperature, $\tilde{\Gamma}_{\mathrm{1PL}}(\omega)$ its mirror function with respect to the ZPL central energy and $n(\omega,T)$ is the Bose-Einstein average number of phonons with frequency $\omega_q = \omega_0 - \omega$.

When inserted in a cavity, the QD interact with an ensemble of  electromagnetic mode whose density is given by the cavity normalized spectrum $S_{\mathrm{cav}}(\omega) = \frac{\kappa}{2\pi}[(\kappa/2)^2+(\omega_c-\omega)^2]^{-1}$, where $\kappa$ is the bare cavity linewidth and $\omega_c$  its frequency. The normalized spontaneous emission spectrum of the QD weakly coupled to a cavity is $S(\omega) = S_{\mathrm{cav}}(\omega)\times S_{\mathrm{QD}}(\omega,T)/\Gamma$ where $\Gamma= \int d\omega\ S_{\mathrm{cav}}(\omega)\ S_{\mathrm{QD}}(\omega,T)$. $S(\omega)$ is  obtained experimentally by collecting the emission through the cavity channel (Fig.1a).  The emission spectrum presents two terms arising from the QD ZPL and  1PL. Since the cavity linewidth is much larger than  the ZPL linewidth, the first  term is given by $S_{\mathrm{cav}}(\omega)\times \Gamma_{\mathrm{ZPL}}(\omega) \approx \Gamma_{\mathrm{ZPL}}(\omega) $. It corresponds to the QD like emission line in the experiment. The second term stems from the QD phonon sideband filtered by the cavity mode $S_{\mathrm{cav}}(\omega)\times \Gamma_{\mathrm{1PL}}(T,\omega)$. It is this term that gives rise to the cavity like emission, whose behavior is now theoretically analysedIt corresponds to the cavity like emission in the experiment.

\begin{figure}[h!]\begin{center}\includegraphics[width=1\linewidth]{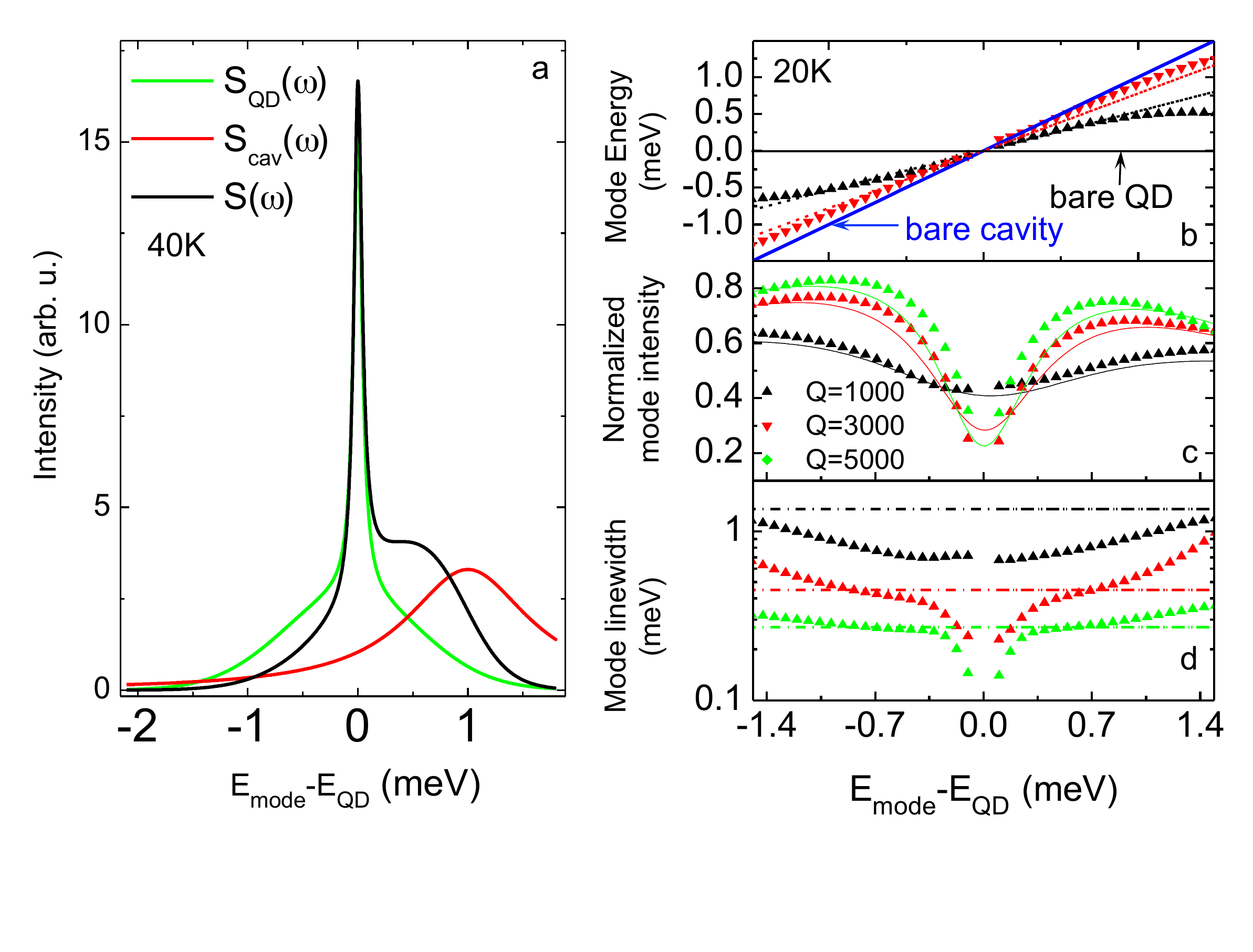}\caption{Cavity pulling induced by the QD phonon bath: a: Calculated emission spectra for a QD-cavity detuning of 1 meV at 40K for a bare cavity of  Q=1000. Green line: calculated emission spectrum $S_{\mathrm{QD}}$ of the QD state coupled to the phonon bath. Red line: cavity spectral function $S_{\mathrm{cav}}$. Black curve: calculated emission spectrum $S$ of the coupled QD-cavity system. b-d: Calculated emission characteristic for the cavity emission line as a function of detuning with the QD line for T=20K. b: Energy of the cavity line with respect to the QD energy. c: Intensity of the cavity line d: Linewidth of the cavity line (log scale). Three bare cavity quality factors are considered: Q=1000 (black), Q=3000 (red), Q=5000 (green). In b, the case of Q=5000 is not shown for clarity and the black (resp. blue) solid line is the bare QD (resp. cavity) frequency. The dotted lines are the linear approximations (equation 1). In d, the horizontal lines show the bare cavity linewidth. Lines in c are the  normalized intensity of the cavity-filtered 1PL:$\int S_{\mathrm{cav}}(\omega)\times \Gamma_{\mathrm{1PL}}(T,\omega) d\omega /\Gamma$ .  }\label{fig:scheme}\end{center}\end{figure}

 Figure 3a shows the calculated spectra for a cavity mode with quality factor Q=1000,  positively detuned by 1 meV from the QD ZPL, at  40 K. The green  line in figure 3a presents the QD emission spectrum $S_{\mathrm{QD}}(\omega,T)$. The main peak corresponds to the ZPL. The phonon sideband  is almost symmetrical on both sides of the ZPL because a significant phonon population. The red line represents the cavity spectrum $S_{\mathrm{cav}}(\omega,T)$. The black curve is the calculated emission spectrum $S$ showing two peaks with a cavity emission line  pulled toward the QD line. Moreover, the cavity emission linewidth is significantly narrower than the bare cavity spectrum.

To elucidate the influence of various parameters on the cavity distortion, $S$ is calculated as a function of the detuning. The apparent cavity energy, intensity and linewidth are extracted from a multi-lorentzian fit to mimic the experimental procedure. Figure 3c presents these parameters as a function of detuning for three bare cavity quality factors $Q=\omega_c/\kappa$. The displaced mode energy is plotted with respect to the QD ZPL. The blue line represents the bare cavity mode. Figure 3b shows that the cavity emission is pulled toward the QD energy.  This pulling is strongest when the mode Q is low. 

 It can be shown that the emission maximum is obtained at a frequency given by 

\begin{equation}
\frac{\kappa \ \omega_{QD}+\kappa_{1PL} \ \omega_c}{\kappa+\kappa_{1PL} }
\end{equation}
approximating the 1PL with a lorentzian shape of linewidth $\kappa_{1PL}$  and considering the small detunings range $\vert \omega_c -\omega_{QD}\vert << \kappa_{1PL}, \kappa$. Note that this simple equation is very similar to the one giving the pulled frequency in atomic gas laser \cite{siegman}. It is plotted in figure 3.b (dotted lines) for  $\kappa_{1PL}=1.5 meV$ and gives a good approximation of the cavity pulling effect for small detunings. In particular, the cavity like behavior follows the bare cavity behavior when the Q is high. This corresponds to a case where phonon assisted mechanisms act as a built in white light source inside the resonator. On the contrary, strong cavity pulling is expected when the Q is low, as observed in Fig. 3a and in Fig.2. 
 
 The symbols in figure 3c show that the mode intensity decreases when approaching the resonance with the QD line, as observed experimentally. The normalized cavity-filtered 1PL  emission $\int S_{\mathrm{cav}}(\omega)\times \Gamma_{\mathrm{1PL}}(T,\omega) d\omega /\Gamma$ is also plotted (thin solid lines). A  good agreement is found with the intensity deduced from the lorentzian fit.
Last, figure 3d shows the calculated cavity linewidth as a function of detuning (symbols). For small quality factors, the phonon sideband mainly leads to a narrowing of the cavity line, i.e. an apparent increase of  Q. For larger Q both apparent reduction or increase of Q are predicted, depending on the slope of the phonon sideband spectrum at a given detuning. 

Our  theoretical framework allows to account for most of the experimental observations as displayed in figure 2.g-i.   The QD-cavity emission properties are calculated  for several QD lines coupled to the cavity mode. To simplify, the QD-cavity detuning is assumed linear with detuning. The measured spectrum is the average of spectra corresponding to the QD being either in the X, CX or XX state. Here, equal occupation probability of the states is assumed.  Each QD line experiences the optimal coupling to the cavity mode and to the phonons. 
As observed in figure 2d-f,  the calculated cavity mode is pulled toward the closest QD transition. In between two transitions, the mode is fed by the phonon sidebands of both  QD lines, resulting in significant slope changes in the line energy and in cavity line broadening.
 Finally, close to each resonance, the increase in the QD emission corresponds to a decrease in the cavity line intensity, as observed experimentally. Indeed, the QD  emits a single photon either in the 1PL or in the ZPL, with a fraction depending on the Purcell effect of each line. When the ZPL is on resonance, the ZPL Purcell enhancement is larger than for the 1PL: the single photon is mainly emitted in the ZPL, at the expense of the 1PL, feeding the cavity line.

While most experimental observations are well accounted theoretically, we note that our experiments show  additional features very close to each QD line resonance where the cavity line presents an avoided crossing with the mode. This mode pushing is especially clear for the CX resonance (see figure 2d). These features are not reproduced theoretically using the bulk phonon dispersion.  Since the cavity probes the phonon sideband emission, the avoiding crossing at resonance  indicates  the presence of discrete phonon modes. The existence of such confined phonon modes has recently been proposed in III-V micropillars cavities \cite{fainstein}.

In conclusion, we have evidenced cavity pulling effects for a single artificial atom coupled to a microcavity in the weak coupling regime. These observations,  obtained thanks to a full control of the emitter-cavity coupling, show a strong distortion of the cavity spectrum induced by the phonon sideband. We note that emission spectroscopy is commonly used to  measure the properties of a cavity mode coupled to a solid state single emitter, whether the emitter is a self assembled QD, a NV center in diamond or a colloidal QD. However, our work shows that this techniques can lead to strong  uncertainties on both the deduced cavity linewidth and spectral position: the coupling to the solid state environment induces new  phenomena appearing in the bad emitter limit for cavity quantum electrodynamics.

\vspace{0.5cm}
\begin{acknowledgments}

{\bf Acknowledgments}

 This work was partially supported by the French ANR P3N DELIGHT, ANR P3N CAFE, the Fondation Nanosciences de Grenoble, the ERC starting grant 277885 QD-CQED, the CHISTERA project SSQN, the National Science Center in Poland (DEC-2011/01/N/ST3/04536 ), and a ``Lider'' grant from The National Center for Research and Development in Poland. The authors thank Dario Gerace, Robson Ferreira, Claude Fabre, Jacqueline Bloch, Thomas Coudreau and Piotr Kossacki for discussions.
\end{acknowledgments}

\end{document}